\documentclass[twoside]{article}
\usepackage{amsfonts,amssymb,amsmath,latexsym}
\usepackage{epsfig,graphicx}
\usepackage{indentfirst}
\usepackage{cite}
\usepackage{color}
\newcommand{\mytitle}[1]{\large \sc #1 \\}
\newcommand{\avtor}[1]  {\large \it #1 \\}

mm  \leftmargin 5pt
\makeatletter \@addtoreset{equation}{section}
\@addtoreset{teo}{section} \@addtoreset{lem}{section}
\@addtoreset{demo}{section} \@addtoreset{rem}{section}
\@addtoreset{prop}{section} \@addtoreset{sled}{theo} \makeatother

\begin{document}
\begin{center}

\mytitle{Berry phases for the nonlocal Gross-Pitaevskii equation
with a quadratic potential}

\bigskip

\avtor {Litvinets$^\#$\footnote{e-mail:
litvinets@mph.phtd.tpu.edu.ru} F.N., Trifonov$^*$\footnote{e-mail:
trifonov@mph.phtd.tpu.edu.ru} A.Yu., and
Shapovalov$^{*\#}$\footnote{e-mail: shpv@phys.tsu.ru} A.V.}

$^{*}$ {\it Mathematical Physics Laboratory\\
Tomsk Polytechnic University,\\
Lenin ave., 30, Tomsk, 634050, Russia}\vskip 0.4cm
$^{\#}$ {\it Theoretical Physics Faculty,\\
Physics Department of
Tomsk State University,\\
Lenin ave., 36, Tomsk, 634050, Russia}
\end{center}

\begin{abstract}

A countable set of asymptotic space -- localized solutions is
constructed by the complex germ method in the adiabatic
approximation for the nonstationary Gross -- Pitaevskii equation
with nonlocal nonlinearity and a quadratic potential. The
asymptotic parameter is $1/T$, where $T\gg1$ is the adiabatic
evolution time.

A generalization of the Berry phase of the linear Schr\"odinger
equation is formulated for the Gross-Pitaevskii equation. For the
solutions constructed, the Berry phases are found in explicit
form.

\end{abstract}

\section*{Introduction}

A quantum system with slowly (adiabatically) varying parameters is
characterized by the conservation of its quantum numbers
throughout the time of adiabatic evolution when the system
Hamiltonian has a nondegenerate spectrum. In other words, the
vector of the system state remains an eigenvector at any time and
gains a phase factor during the adiabatic evolution (see, e. g.,
\cite{Messia}). M.~Berry has shown \cite{shapovalov:BERRY}) that
the phase gained  consists of a dynamic and a geometric
(topological) part. Such a division is caused by different
mechanisms of their origination. The dynamic phase is associated
with the  evolution of the system and the topological one with the
geometry of its parameter space.

Both phases result from  the decomposition of the general phase of
the leading term in the  semiclassical approximation of the
solution of the Schr\"odinger equation in a small parameter $1/T$,
where $T$ is the adiabatic evolution time. The dynamic phase
corresponds to the zeroth order term in the $1/T$-expansion, and
the Berry phase belongs to the first order term. Inclusion of the
topological phase in the general phase is necessary to determine
the leading term of the semiclassical asymptotic series. The work
\cite{shapovalov:BERRY} has initiated comprehensive investigations
of the geometric phases of linear quantum-mechanical equations.
For details see the reviews
\cite{shapovalov:VINI,Moore,shapovalov:KLYSHKO,Biswas}.

It is of interest to study geometric phases for nonlinear
equations describing various class of nonlinear phenomena
\cite{Garrison} whose nonlinear properties are due to  collective
interactions.  An important example of systems showing nonlinear
properties is the Bose-Einstein condensate (BEC)\cite{
shapovalov:PITAEVSKII, shapovalov:PITAEVSKII-2}.

In the BEC theory various models are used. In
\cite{Fazakon1,Fazakon3} the BEC model is based on the two-level
Hamiltonian of  a system of particles confined by a magnetic
field. The system has a constant number of particles distributed
between two subsystems, which are in different states. The Berry
phase for this system is obtained under the condition that the
system parameters responsible for the interaction of the
subsystems vary slowly.

Many of the BEC models are based on the Gross-Pitaevskii equation
(GPE) \cite{shapovalov:PITAEVSKII-1, shapovalov:GROSS,
shapovalov:PITAEVSKII, shapovalov:PITAEVSKII-2}, which is a
non-stationary multidimensional cubic-nonlinear Schr\"odinger
equation with the potential  of an external field. The BEC states
are described by localized solutions of the GPE. In studying these
models, a number of serious  mathematical problems arise. We
mention two of them which are essential for our consideration.

First, no method of exact integration  for the GPE are available
except for the nonstationary one-dimensional Schr\"odinger
equation that is integrable by the Inverse Scattering Transform
method \cite{ZAKHAROV} with no external field. In the presence of
an external field, only an approximate solution can be found in
terms of the soliton perturbation theory \cite{kivshar-malomed}
with the assumption of a weak external field.

Second, even for a two-dimensional space, the localized solutions
of the nonlinear Schr\"odinger equation (NLSE) with focusing cubic
nonlinearity and no external field are unstable and eventually
collapse \cite{BANG}, which is not  observed in experiments. On
the other hand, models based on the GPE with local nonlinearity
can be considered as simplified versions of models with the GPE
having a nonlocal nonlinearity \cite{Deconinck}. This is a reason
to consider in more detail the nonlocal operator
$\int\limits_{-\infty}^{+\infty}\, {V(\vec{x}-\vec{y})
|\Psi(\vec{y},t)|^2}d\vec{y}$   that arises in the derivation of
the Gross-Pitaevkii equation. The interatomic potential
$V(\vec{x}-\vec{y})$ is usually assumed to be short-range, and
therefore the above nonlocal nonlinear operator can be replaced by
a local operator $\beta |\Psi(\vec{x},t)|^2$, where
$\beta=\int\limits_{-\infty}^{+\infty}\,{V(\vec{y})}d\vec{y}$.
More detailed analysis of the nonlocal properties can be found,
for example, in \cite{Deconinck} where the wave function of the
one-dimensional equation is expanded in a series in $z=x-y$. The
symmetry of the potential and its decrease at infinity result in
an additional term $\displaystyle\propto \frac{\partial^{2}
}{\partial x^{2}}\mid\psi\mid^2$ in the local Gross-Pitaevskii
equation. The equations obtained are studied by numerical methods
as  the construction of an analytical solution fails even in the
one-dimensional case.

In this work we consider the one-dimensional nonlocal
Cross-Pitaevskii equation and use an approach where the potential
$V(x,y)$ is expanded in the variables $(x,y)$ to the second-order
terms. Exact solutions are constructed for this equation in
\cite{Litv:LTS1,Litv:LTS3}. Although the interatomic quadratic
potential does not decrease at infinity, the
 convergence of the integral is provided by a proper choice of the class of functions in
which we seek solutions to the equation.

The one-dimensional GPE  describes  BEC states if the longitudinal
dimension of the condensate is much greater than its cross
dimension \cite{Deconinck}. Unlike the NLSE with local
nonlinearity, the  basic properties of the solutions of the
nonlocal one-dimensional GPE conserve in the multidimensional
case.

The aim of this work is to obtain the Berry phase
in explicit form for a one-dimensional  GPE  of the form
\begin{eqnarray}
&&\bigg\{ -i\hbar\partial_t
+\widehat{\mathcal H}_{\varkappa}(R(t),\Psi(t))\bigg\}\Psi =0,\label{pgau1}\\
&&\widehat{\mathcal H}_{\varkappa}(R(t),\Psi(t))=\widehat{\mathcal
H}(R(t))+\varkappa\widehat V(R(t),\Psi(t)),\label{phbb07}\\
&&\displaystyle\widehat{\mathcal H}(R(t))= \frac{\mu(t)
 p^2}{2} +\frac{\sigma(t) x^2}{2}+\frac{\rho(t)(x\hat p+\hat px)}2,\cr
&&\displaystyle\widehat V(R(t),\Psi(t))=\frac12
  \int\limits_{-\infty}^{+\infty}dy\,
{\Big[a(t)x^2+2b(t)xy+c(t)y^2\Big] |\Psi(y,t)|^2}\nonumber
\end{eqnarray}
which involves a quadratic potential and an external field of
the harmonic oscillator type. Here $a(t)$, $b(t)$,  and $c(t)$ are the
potential parameters; $\varkappa$ is the nonlinearity parameter; and
$\mu(t)$, $\sigma(t)$, and $\rho(t)$ are time-dependent system
parameters. Therefore, the Hamiltonian depends on time via
the set of parameters $R(t)$=($\mu(t)$, $\sigma(t)$, $\rho(t)$,
$a(t)$, $b(t)$, and $c(t)$).
The quadratic potential in Eqs.
(\ref{pgau1}), (\ref{phbb07}) models the magnetic traps that confine the
condensate \cite{Shchesnovich}.

The  problem of constructing solutions for Eq. (\ref{pgau1}) with
an external potential is solved with the use of  the semiclassical
integration method developed in \cite{Litv:LTS1,Litv:LTS3,
shapovalov:BTS1,shapovalov:BTS2}, where Eq. (\ref{pgau1}) is
called the "Hartree type equation".  Below we call the nonlinear
operator $\widehat{\mathcal H}_{\varkappa}$ in Eq.  (\ref{pgau1})
the nonlinear Hamiltonian. The semiclassical integration method
allows one to construct localized solutions, approximate for the
potential of general form and exact for the quadratic potential.
We use these solutions  to find and study geometric Berry phases.

Let us give necessary facts from the theory of geometric phases of
linear quantum mechanical systems to find Berry  phase in the
nonlinear case. Following \cite{shapovalov:BERRY} (see also
\cite{shapovalov:KLYSHKO,shapovalov:VINI}), we consider a
Hamiltonian $\widehat{\mathcal H}(t)=\widehat{\mathcal H}(R(t))$,
which depends on time $t$ via a set of slowly varying $T$-periodic
functions $R(t)$. Denote by \, $\Psi_{E_n(R(t))}(x,R(t))$ the
eigenfunctions of the instantaneous  Hamiltonian
$\widehat{\mathcal H}(R(t))$:
\begin{equation}
\widehat{\mathcal H}(R(t))\Psi_{E_n(R(t))}(x,R(t))=
E_n(R(t))\Psi_{E_n(R(t))}(x,R(t)).\label{Ber2_2}
\end{equation}
Assume the spectrum of the Hamiltonian  $\widehat{\mathcal
H}(R(t))$ to be nondegenerate at any fixed time and  set the
Cauchy problem
\begin{equation}
\big\{ -i\hbar\partial_t +\widehat{\mathcal H}(R(t))\big\}\Psi
=0,\label{gau1}
\end{equation}
\begin{equation}
\Psi\big|_{t=t_0}=\Psi_{E_n(R(t_0))}(x,R(t_0)).\label{Ber2_3}
\end{equation}
According to the adiabatic theorem \cite{Messia},\footnote{It is
generally agreed that a system evolves adiabatically if
\begin{equation}
\max_{i=\overline{1,n}}\dot{R_{i}}\frac{T}{R_{i}}\ll
1,\label{phb07}
\end{equation}
where $R_{i}$ are the  parameters of the Hamiltonian (see
\cite{Landau}).} the quantum numbers of the system will conserve
throughout the adiabatic evolution time $T$ if the parameters
$R(t)$ depend on time adiabatically. In other words, a  solution
of the Cauchy problem (\ref{gau1}), (\ref{Ber2_3}) differs from
the initial state by the phase factor
\begin{equation}
\Psi(x, T)=\exp[{i}\phi_{n}(T)]\Psi_{E_n(R(T))}(x,R(T)).
\label{Ber2_4}
\end{equation}

Following \cite{shapovalov:BERRY}, represent the phase
$\phi_{n}(T)$ as
\begin{equation}
\phi_{n}(T)=\delta_{n}(T)+\gamma_{n}(T), \label{Ber2_5}
\end{equation}
where  $\delta_{n}(T)$ is the dynamic phase given by
\begin{equation}
\delta_{n}(T)=-\frac{1}{\hbar}\int\limits_{0}^{T}E_n(R(t))dt.\label{phb01}
\end{equation}
The phase $\gamma_{n}(T)$ is called an {\it adiabatic Berry
phase}, and for a linear Schr\"odinger equation it is determined
by the expression
\begin{equation}
\dot{\gamma}_{n}(t)=i\langle\Psi_{E_n(R(t))}(x,R(t))|
\dot{\Psi}_{E_n(R(t))}(x,R(t))\rangle\label{phb02}
\end{equation}
or
\begin{equation}
\gamma_{n}(T)=i\int\limits_{0}^{T}\langle\Psi_{E_n(R(t))}(x,R(t))|
\dot{\Psi}_{E_n(R(t))}(x,R(t))\rangle
dt=\oint\limits_{C}A^{n}_{j}dR^{j}.\label{phb04}
\end{equation}
Here
\begin{equation}
A^{n}_{j}=i\Big\langle\Psi_{n}\mid\frac{\partial\Psi_{n}}{\partial
R^{j}}\Big\rangle,\label{phb05}
\end{equation}
and $C$ is a closed contour in the parameter space. The functions
$ A^{n}_{j}$ act as components of an  "induced gauge field" since
under the gauge transformations
\begin{equation}
\Psi_{n}\to\exp(i\xi_{n}(R))\Psi_{n}
\end{equation}
the quantities $A^{n}$ are transformed as
\begin{equation}
A^{n}_{j}\to A^{n}_{j}+\frac{\partial\xi_{n}}{\partial
R^{j}}.\label{qphb05}
\end{equation}
Expression  (\ref{phb04}) does  not depend on the transformation
(\ref{qphb05}) as the contour $C$ is closed. The dynamic phase
$\delta_{n}(T)$ characterizes the mean value of the system energy,
and the geometric phase $\gamma_{n}(T)$ does not depend on the
system dynamics. Berry's phase depends on  the geometry of the
system parameter space and on the type of the contour $C$.

Assume the adiabatic theorem to be hold for nonlinear equations
and  define Berry's phase by  relations (\ref{Ber2_5}) and
(\ref{phb01}). Formula  (\ref{phb04}), being equivalent to
(\ref{Ber2_5}) in the linear case,  needs additional
substantiation for nonlinear equations.

In this work we use the approach developed in \cite{DodKlMan,ty94}
for linear equations to find Berry's phase. The approach is based
on the exact (or approximate) solution of the Cauchy problem
(\ref{gau1}), (\ref{Ber2_3}) which is expanded in an adiabatic
parameter. For the linear equation (\ref{pgau1}) the exact
solution of the Cauchy problem is constructed using the method
proposed in \cite{Litv:LTS1}. Here, the initial Hamiltonian  in
the initial condition of the Cauchy problem (\ref{Ber2_3}) is
replaced by (\ref{phbb07}) and the solution is determined by  two
auxiliary systems of ordinary differential equations: the
Hamilton-Ehrenfest system and the system in variations. Solutions
of these systems are unknown when the coefficients are arbitrary
functions of time. When the coefficients depend on time
adiabatically, we can seek  the solution in the form of an
expansion in the adiabaticity parameter which is taken as $1/T$,
where  $T$ is a  "long" characteristic time, for example, the
adiabatic evolution period of the system.

If the Hamilton-Ehrenfest system and the system in variations are
solved accurate to $O(1/T)$,  the equation considered  is solved
with the same  accuracy, and thus we obtain a solution of equation
(\ref{pgau1}) in the adiabatic approximation. For such a solution,
Berry's phase can be found in explicit form.

\section{The Hamilton-Ehrenfest system}

For a linear operator $\widehat A$ we define its  mean value in a
state $\Psi(t)$ as
\begin{equation}
\label{gau1a} \langle\widehat A(t)\rangle=\frac{1}{\|\Psi(t)\|^2}
\langle\Psi(t)|\widehat A| \Psi(t) \rangle=A_\Psi(t,\hbar).
\end{equation}
On the solutions $\Psi(t)$ of equation  (\ref{gau1}) we have
\begin{eqnarray}
\label{gau1b} &\displaystyle\frac{d\langle\widehat A(t)\rangle}{d
t}=\Big\langle\frac{\partial \hat A(t)}{\partial t}
\Big\rangle+\frac{i}{\hbar}\langle [\widehat{\mathcal
H}_\varkappa(t,\Psi(t)),\widehat A(t)] \rangle,
\end{eqnarray}
where $[\hat A,\hat B]=\hat A\hat B-\hat B\hat A$ is the
commutator of linear operators  $\hat A$ and  $\hat B$.

Similar to the linear case, we call  (\ref{gau1b}) the Ehrenfest
equation. From this equation with $\widehat A=1$ it follows, in
particular, that the norm of  a solution of equation (\ref{gau1})
is conserved, that is,
\[
\|\Psi(x,t)\|^2=\|\Psi(x,0)\|^2=\|\Psi\|^2.
\]
So it is convenient to use the nonlinear parameter
$\tilde\varkappa=\varkappa\|\Psi\|^2$ instead of $\varkappa$.

Denote by
\[ \alpha_\Psi^{(l,k)}(t,\hbar)=\frac{1}{\|\Psi\|^2}
\int\limits_{-\infty}^{+\infty}\Psi^*(y,t)\{(\Delta \hat p_y)^{l}
(\Delta y)^{k}\}\Psi(y,t)dy, \quad k,l=\overline{0,\infty},
\]
the moments  of the $(k+l)$ - th order centered relative to
$x_\Psi(t,\hbar)$ and  $p_\Psi(t,\hbar)$. Here  $\Delta \hat p_y
=-i\hbar\partial_y-p_\Psi(t,\hbar)$, and  $\{(\Delta \hat
p_y)^{l}(\Delta y)^{k}\}$ is a Weyl-ordered operator with the
symbol $(\Delta p_y)^{l}(\Delta y)^{k}$. In particular,
\begin{eqnarray*}
&& \sigma_{xx}(t,\hbar)=\alpha_\Psi^{(0,2)}(t,\hbar), \\[6 pt]
&& \sigma_{pp}(t,\hbar)=\alpha_\Psi^{(2,0)}(t,\hbar),\\[6 pt]
&& \sigma_{xp}(t,\hbar)=\alpha_\Psi^{(1,1)}(t,\hbar)
\end{eqnarray*}
are the variances of the coordinates and momenta  and of the
correlation function of the coordinates and momenta, respectively.

Consider the first-order operators $\hat p$ and $x$ and the
centered second-order operators  $(\Delta x)^2$, $(\Delta \hat
p)^2$,\linebreak $(\Delta x\Delta \hat p+\Delta \hat p\Delta
x)/2$, where $\Delta \hat p=\hat p-p_\Psi(t, \hbar)$, $\Delta
x=x-x_\Psi(t,\hbar)$, $p_\Psi(t,\hbar)=\langle \hat p\rangle$, and
$x_\Psi(t, \hbar)=\langle x\rangle$.

The Ehrenfest system for the mean values of these operators reads
\begin{equation}
\left\{\begin{array}{l} \dot{p}=-\sigma_{0}(t)x-\rho(t)p,\\[4pt]
\dot{x} =\mu(t) p+\rho(t) x,\\[4pt]
\dot{\sigma}_{xx}=2\mu(t) \sigma_{xp}+2\rho(t)\sigma_{xx}, \\[4pt]
\dot{\sigma}_{xp}=\mu(t)\sigma_{pp}-\tilde{\sigma}(t)\sigma_{xx},\\[4pt]
\dot{\sigma}_{pp}=-2\rho(t)\sigma_{pp}-2\tilde{\sigma}(t)\sigma_{xp}.
\end{array}\right.\label{gau3n}
\end{equation}
Here
\begin{eqnarray*}
 \sigma_{0}(t)=\sigma(t)+\tilde\varkappa(a(t)+b(t)),\quad
 \tilde{\sigma}(t)=\sigma(t)+\tilde\varkappa a(t).
\end{eqnarray*}
For system  (\ref{gau3n}), let us set a Cauchy problem with
initial conditions
\begin{equation}\begin{array}{l}
p|_{t=s}=p_0 \qquad
x|_{t=s}=x_0 ,\\[6 pt]
{\sigma}_{pp}|_{t=s}={\sigma}_{0pp},
\quad{\sigma}_{xp}|_{t=s}={\sigma}_{0xp},
\quad{\sigma}_{xx}|_{t=s}={\sigma}_{0xx}.
\end{array}\label{llst03a3a}
\end{equation}
We call (\ref{gau3n}) {\em the second order Hamilton-Ehrenfest
system } (HES)  related to equation (\ref{gau1}).

Consider the HES (\ref{gau3n}) as a dynamical system which is not
related to equation (\ref{gau1}). Apparently, not all solutions of
the HES can be obtained as mean values of the corresponding
operators on the solutions of equation (\ref{gau1}). For example,
the mean values must satisfy the Schr\"odinger uncertainty
relation
\begin{equation}
\sigma_{pp}\sigma_{xx}-\sigma_{xp}^2\ge\frac{\hbar^2}4\label{llst03a4a}
\end{equation}
for the second-order moments (for the higher-order relations see
\cite{Robertson}). It can readily be seen that the HES admits the
trivial solution  $p=0$, $x=0$, $\alpha^{(k,l)}=0$, $k+l=2$. The
left-hand side of (\ref{llst03a4a}) is the integral of motion of
the HES (\ref{gau3n}) (see \cite{bk1}). Hence, it suffices that
the uncertainty relation be fulfilled at the initial time. The
uncertainty relations will be fulfilled automatically if the
initial conditions for (\ref{gau3n}) are taken as
\begin{equation}\begin{array}{l}
p|_{t=s}=p_0= p_\psi(\hbar), \qquad x|_{t=s}=x_0= x_\psi(\hbar) ,\\[6 pt]
{\sigma}_{pp}|_{t=s}=\alpha_{\psi}^{(2,0)}(\hbar),
\quad{\sigma}_{xp}|_{t=s}=\alpha_{\psi}^{(1,1)}(\hbar),
\quad{\sigma}_{xx}|_{t=s}=\alpha_{\psi}^{(0,2)}(\hbar),
\end{array}\label{llst03a3aq}
\end{equation}
where $\psi(x,\hbar)$ is the initial condition for equation
(\ref{gau1}):
\begin{equation}
\Psi(x,\hbar,t)\big|_{t=0}=\psi(x,\hbar).\label{gau3b}
\end{equation}
Denote  a trajectory in an extended phase space by ${\mathfrak
g}={\mathfrak g}(t,{\mathfrak C})\in{\mathbb R}^5$, where
\begin{eqnarray}
&&{\mathfrak g}(t,{\mathfrak C})=\big(P(t,{\mathfrak
C}),X(t,{\mathfrak C}),\sigma_{pp}(t,{\mathfrak C}),
\sigma_{px}(t,{\mathfrak C}),\sigma_{xx}(t,{\mathfrak C})\big)^\intercal, \label{llst03a30}\\
&& {\mathfrak C}=(C_1,C_2,C_3,C_4,C_5)^\intercal\nonumber
\end{eqnarray}
is the general solution of the Hamilton-Ehrenfest system
(\ref{gau3n}) and $\hat{\mathfrak g}$ is the operator column
\begin{equation}
\hat{\mathfrak g}=\Big(\hat p, \hat x,(\Delta\hat p)^2,\frac
12(\Delta\hat p\Delta x -\Delta x\Delta\hat p),(\Delta
x)^2\Big)^\intercal.\label{llst03a30a}
\end{equation}
Here  $C_l$, $l=\overline{1,5}$ are arbitrary constants which can
be expressed in terms of the initial conditions (\ref{llst03a3a}).
The matrix $B^\intercal$ is transposed to the matrix $B$. The
system (\ref{gau3n}) can be rewritten in the form
\begin{equation}
\dot{\mathfrak g}={\mathfrak A} {\mathfrak g}, \qquad {\mathfrak
g}\big|_{t=s}={\mathfrak g}_0, \label{llst03a30b}
\end{equation}
where
\[{\mathfrak A}=\left(\begin{array}{ccccc}
-\rho(t)&-\sigma_{0}(t)&0&0&0\\[4pt]
\mu(t)&\rho(t)&0&0&0\\[4pt]
0&0&-2\rho(t)&-2\tilde{\sigma}(t)&0\\[4pt]
0&0&\mu(t)&0&-\tilde{\sigma}(t)\\[4pt]
0&0&0&2\mu(t)&2\rho(t)
\end{array}\right).\]

\section{The associated linear  Schr\"odinger equation}

Let us seek a solution to equation (\ref{pgau1}) in the form of
the ansatz
\begin{equation}
\Psi(x,t,\hbar)=\varphi\Bigl(\frac{\Delta x}{\sqrt{\hbar}},t,
  \sqrt{\hbar}\Bigr)\exp\Bigl[{\frac{i}{\hbar}
  \Bigl(S(t,{\mathfrak C})+P(t,{\mathfrak C}) \Delta x \Bigr)}\Bigr].
  \label{gau4}
\end{equation}
Here the function  $\varphi(\xi ,t,\sqrt{\hbar} )$ belongs to the
Schwartz space ${\Bbb S}$ in the variable $\xi=\Delta
x/\sqrt\hbar$  and depends regularly  on $\sqrt{\hbar}$; $\Delta
x=x-X(t,{\mathfrak C})$. The real functions  $S(t,{\mathfrak C})$
and $Z(t,{\mathfrak C})=(P(t,{\mathfrak C}),X(t,{\mathfrak C}))$
that characterize the solution are to be determined.

Expand the operators in equation (\ref{pgau1}) in a Tailor series
in
 $\Delta x=x-x_\Psi(t,\hbar)$, $\Delta
y=y-x_\Psi(t,\hbar)$, and  $\Delta \hat p=\hat p-p_\Psi(t,\hbar)$.
Then equation (\ref{pgau1}) takes the form
\begin{eqnarray}
&&\displaystyle\{ -i\hbar\partial_t +{\mathfrak H}(t,\Psi)+
\langle{\mathfrak H}_z(t,\Psi),\Delta\hat z\rangle
+\frac12\langle\Delta\hat z,{\mathfrak H}_{zz}(t,\Psi)\Delta\hat z\rangle\}\Psi=0, \label{gau6}\\
&&\displaystyle{\mathfrak H}(t,\Psi)=\frac{\mu(t)
p_\Psi^2(t,\hbar)}{2}+\frac{\sigma(t) x_\Psi^2(t,\hbar)}{2}
+\rho(t) x_\Psi(t,\hbar)p_\Psi(t,\hbar)+\nonumber\\
&&\quad+\displaystyle\frac{\tilde\varkappa}2
c\alpha_\Psi^{(0,2)}(t,\hbar) +\frac{\tilde\varkappa}2(a+2b+c)
x_\Psi^2(t,\hbar), \nonumber\\
&&{\mathfrak H}_z(t,\Psi)= \left(
 \begin{array}{c}\mu(t) {p_\Psi(t,\hbar)}+\rho(t) x_\Psi(t,\hbar)\\[4pt]
 \sigma(t) x_\Psi(t,\hbar)+\rho(t)
 p_\Psi(t,\hbar)+\tilde\varkappa(a+b)
 x_\Psi(t,\hbar)\end{array}\right),\nonumber\\
&& {\mathfrak H}_{zz}(t,\Psi)=\left(\begin{array}{cc}
   \mu(t) &\rho(t) \\[4pt]
   \rho(t) & \widetilde{\sigma}(t)\end{array}\right). \nonumber
\end{eqnarray}

Let us associate the nonlinear equation  (\ref{gau6}) with the
linear equation that is obtained from (\ref{gau6}) by formal
substitution of the solution of the HES (\ref{gau3n}) instead of
the corresponding mean values of the coordinate and momenta
operators and  second-order centered moments. The resulting linear
equation is
\begin{eqnarray}
&&\displaystyle\{-i\hbar\partial_t +{\mathfrak H}(t,{\mathfrak
 C})+\langle{\mathfrak H}_z(t,{\mathfrak C}),\Delta\hat z\rangle
 +\frac12\langle\Delta\hat z,{\mathfrak H}_{zz}(t,{\mathfrak C})
 \Delta\hat z\rangle\}\Phi=0, \label{gau6a}\\
&&\displaystyle{\mathfrak H}(t,{\mathfrak C})=\frac{\mu(t)
 P^2(t,{\mathfrak C})}{2}+\frac{\sigma(t) X^2(t,{\mathfrak
 C})}{2}+\rho(t) X(t,{\mathfrak C})P(t,{\mathfrak C})+\nonumber\\
&&\quad+\displaystyle\frac{\tilde\varkappa}2
 c\sigma(t)_{xx}(t,{\mathfrak C},\hbar)
 +\frac{\tilde\varkappa}2(a+2b+c) X^2(t,{\mathfrak C}),\nonumber\\
&&{\mathfrak H}_z(t,{\mathfrak C})= \left(\begin{array}{c}
 \mu(t) P(t,{\mathfrak C})+\rho(t) X(t,{\mathfrak C})\\[4pt]
 \sigma_0(t) X(t,{\mathfrak C})+\rho(t)
 P(t,{\mathfrak C})\end{array}\right),\nonumber\\
&& {\mathfrak H}_{zz}(t,{\mathfrak C})=\left(\begin{array}{cc}
 \mu(t) &\rho(t) \\[4pt]
 \rho(t) & \widetilde{\sigma}(t)\end{array}\right).\nonumber
\end{eqnarray}
We call Eq.  (\ref{gau6a}) the  {\em associated linear
Schr\"odinger equation}.

By direct check we see that the function
\begin{equation}
\Phi_0(x,t,{\mathfrak C})=\mid0,t,{\mathfrak
C}\rangle=N_\hbar\Bigl(\frac{C(0)}{C(t)}\Bigr)^{1/2}\exp{
\biggl\{\frac{i}{\hbar}\Big(S(t,{\mathfrak C})+ P(t,{\mathfrak
C})\Delta x+\frac 12\frac{B(t)} {C(t)}\Delta x^2\Big)\biggr\}}
\label{spgau13a}
\end{equation}
is a solution of equation  (\ref{gau6a}). Here
\begin{equation}
S(t,{\mathfrak C})=\int\limits_0^t\big(P(t,{\mathfrak C})\dot
X(t,{\mathfrak C})-{\mathfrak H}(t,{\mathfrak C})\big)dt;
\label{gau13c}
\end{equation}
and  $B(t)$ and $C(t)$ denote, respectively, the momentum and the
coordinate part of the solution
\begin{equation}
a(t)=\left(\begin{array}{l} B(t)\\[4pt]
C(t)\end{array}\right),\label{gau13b}
\end{equation}
of the system in variations
\begin{equation}
\dot a=J{\mathfrak H}_{zz}(t)a, \label{gau13q}\qquad
a\mid_{t=s}=a_0,
\end{equation}
related to equation (\ref{gau6a}).

The normalizing condition  $\|\Psi\|^2=1$ yields
$N_\hbar=(\pi\hbar)^{-1/4}(\mid C(0)\mid)^{-1/2}$.

Let us introduce the notation
\[ \hat a(t)=N_a\big(C(t)\Delta\hat p-B(t)\Delta x\big). \]
If $C(t)$ and  $B(t)$ are solutions of equations (\ref{gau13q}),
then the operator $\hat a(t)$ commutes with the operator of the
associated equation
 (\ref{gau6a}). Therefore, the function
\[
\Phi_n(x,t,{\mathfrak C})=\frac{1}{\sqrt{n!}}\Big(\hat
a^+(t)\Big)^n\Phi_0(x,t,{\mathfrak C}),\qquad
n=\overline{0,\infty},
\]
will be a solution of the Schr\"odinger equation (\ref{gau6a}).
Commutating the operators $\hat a^+(t)$ with the operator of
multiplication by the function
 $\Phi_0(x,t,{\mathfrak C})=\mid 0,t,{\mathfrak C}\rangle$, we obtain the following
 representation for the Fock basis of solutions of the linear equation  (\ref{gau6a}):
\begin{eqnarray*}
&&\Phi_n(x,t,{\mathfrak C})= \displaystyle\frac{(i)^n}{\sqrt{n!}}
\Big( N^*_a\Big)^n\Phi_0(x,t,{\mathfrak C})\Big[\frac{\mid
C(t)\mid
 \sqrt{\hbar}}{C(t)}\Big]^n 
 \Big[\sqrt{\hbar}\mid C(t)\mid\frac\partial{\partial x}-
 \frac{2}{\sqrt{\hbar}|C(t)|}\Delta x\Big]^{n} 1=\\
&&\quad
 =\displaystyle\frac{1}{\sqrt{n!}}\Big( N^*_a\Big)^n\Phi_0(x,t,{\mathfrak C})
 \bigg(\frac{i}{\sqrt{2}}\bigg)^n (\sqrt{\hbar})^n\exp{(-in\, {\rm
 Arg}\, C(t))}H_n \biggl(\Delta x\sqrt{\frac{{\rm Im}\,
 Q(t)}{\hbar}}\biggr),
\end{eqnarray*}
where  $H_n(\xi)$ are the Hermitian polynomials and
$Q(t)=B(t)C^{-1}(t)$. Finding
$N_a\!=\!\Big(\!1/\sqrt{2\hbar}\!\Big)\exp{[-i {\rm Arg}\, C(0)]}$
from the condition $[\hat a(t),\hat a^+(t)]=1$, we have
\begin{eqnarray}
&&\Phi_n(x,t,{\mathfrak C})=\mid n,t,{\mathfrak C}\rangle=
 \displaystyle\frac{1}{n!}[\hat{a^{+}}(t)]^{n}\mid0,t,{\mathfrak
 C}\rangle=\cr
&&\quad=\displaystyle\frac{1}{\sqrt{n!}}\mid0,t\rangle
 \bigg(\frac{i}{\sqrt{2}}\bigg)^{n}\exp{[-in
 ({\rm Arg}\, C(t)-{\rm Arg}\, C(0))]}H_{n}(\xi),\cr
&&\xi=\sqrt{\displaystyle\frac{{\rm Im}\, Q(t)}{\hbar}}\Delta x.
\label{gau19}
\end{eqnarray}
Using the properties of  Hermitian polynomials, we obtain the mean
values of the momentum and coordinate operators and the
corresponding variances:
\begin{eqnarray}
&& x_{\Phi_n}=0,\qquad p_{\Phi_n}=0,\nonumber\\
&& \alpha^{(2,0)}_{\Phi_n}=\sigma_{pp}(t,\hbar)=
 \hbar\displaystyle\frac{\widetilde{\sigma}(t)(2n+1)}{2\mu(t) {\rm Im}\, Q(t)};\nonumber\\
&& \alpha^{(1,1)}_{\Phi_n}=\sigma_{xp}(t,\hbar)=
 \hbar\displaystyle\frac{\rho(t)(2n+1)}{2\mu(t) {\rm Im}\, Q(t)};\label{gau16c}\\
&& \alpha^{(0,2)}_{\Phi_n}(t,\hbar)= \sigma_{xx}(t,\hbar)=
 \hbar\displaystyle\frac{(2n+1)}{2{\rm Im}\, Q(t)}.\nonumber
\end{eqnarray}

The functions $\Phi_n(x,t,{\mathfrak C})$ are solutions of
equation (\ref{pgau1}) for properly chosen ${\mathfrak C}$, such
that the solutions of the Hamilton-Ehrenfest system solutions
(\ref{gau3n}) coincide with equations  (\ref{gau16c}). Denoting
this set of parameters by $\overline{{\mathfrak C}}_n$, we obtain
\begin{equation}
\Psi_{n}(x,t)=\Phi_n(x,t,\overline{{\mathfrak C}}_n).
\end{equation}
The subscript   $n$ in  $\overline{{\mathfrak C}}_n$ implies that
every function $\Psi_{n}(x,t)$ has its own set of parameters
$\overline{{\mathfrak C}}_n$.


\section{Eigenfunctions of  the instantaneous nonlinear Hamiltonian
}

To construct the Berry phase   using formulas (\ref{Ber2_4}) and
(\ref{Ber2_5}), we have to solve a spectral problem for the
instantaneous Hamiltonian (\ref{phbb07}) in the class of functions
(\ref{gau4})
\begin{equation}
\widehat{\mathcal H}_{\varkappa}(R,\psi_n)\psi_{n} =
E_{n}\psi_{n}.\label{spgau1a}
\end{equation}
To solve this problem, consider the nonstationary Schr\"odinger
equation
\begin{equation}
\big\{ -i\hbar\partial_t +\hat{\mathcal
H}_{\varkappa}\big(R,\Psi(t)\big)\big\}\Psi =0.\label{specgau1}
\end{equation}
Solutions of equation  (\ref{specgau1}) of the form
\begin{equation}
\Psi(x,t)=\exp\Big\{-\frac{i}{\hbar}E_n(R) t\Big\}\psi_n(x,R)
\end{equation}
provide a solution of the spectral problem (\ref{spgau1a}) where
$\psi_n(x,R)$ and $E_n(R)$ are the instantaneous eigenfunctions
and eigenvalues of the Hamiltonian $\widehat{\mathcal
H}_{\varkappa}(R,\psi_n(R))$, respectively.

The Hamilton--Ehrenfest system  (\ref{gau3n}) for the first-order
moments related to equation (\ref{specgau1}) takes the form
\begin{equation}
\left\{\begin{array}{l} \dot{p}=-\sigma_0x-\rho p,\\[4pt]
\dot{x}=\mu p+\rho x,
\end{array}\right.\label{spgau2n}
\end{equation}
and for the second-order moments we have
\begin{equation}
\left\{\begin{array}{l}
\dot{\sigma}_{xx}=2\mu\sigma_{xp}+2\rho \sigma_{xx}, \\[4pt]
\dot{\sigma}_{xp}=\mu\sigma_{pp}-\tilde{\sigma}\sigma_{xx},\\[4pt]
\dot{\sigma}_{pp}=-2\rho\sigma_{pp}-2\tilde{\sigma}\sigma_{xp},
\end{array}\right.\label{spgau3n}
\end{equation}
where
\[
\sigma_{0}=\sigma+\tilde\varkappa(a+b),\quad
\tilde{\sigma}=\sigma+\tilde\varkappa a.
\]
Let us denote
\begin{equation}
\tilde\Omega=\sqrt{\sigma_{0}\mu-\rho^2}, \qquad
\Omega=\sqrt{\tilde{\sigma}\mu-\rho^2}. \label{gau4a}
\end{equation}
The spectral problem is associated only with the time-localized
solutions of the system  (\ref{spgau2n}),  (\ref{spgau3n}) which
are stable in the linear approximation. The localization condition
 holds when
\[
\tilde\Omega^2=\sigma_0\mu-\rho^2>0,\quad\Omega^2=\tilde{\sigma}\mu-\rho^2>0.
\]
In this case the general solution of the system (\ref{spgau2n}) is
given by
\begin{eqnarray}
&& X(t)=C_1\sin\tilde\Omega t+ C_2\cos\tilde\Omega t,
\nonumber \\
&&\displaystyle P(t)=\frac{1}{\mu}\Big(\tilde\Omega C_1-\rho
C_2\Big)\cos\tilde\Omega t -\frac{1}{\mu}\Big(\tilde\Omega C_2
+\rho C_1\Big)\sin\tilde\Omega t. \label{gau15a}
\end{eqnarray}
Accordingly, for the system (\ref{spgau3n}) we have
\begin{eqnarray}
&&\sigma_{xx}(t)=C_3\sin2\Omega t+C_4\cos2\Omega t+C_5,\nonumber \\
&&\sigma_{xp}(t)=\displaystyle\frac{1}{\mu}\bigg(\Omega C_3-\rho
C_4\bigg)\cos2\Omega t -\frac{1}{\mu}\Big(\Omega C_4+\rho C_3\Big)
\sin2\Omega t-\frac{\rho}{\mu} C_5,\label{gau15d}\\
&&\sigma_{pp}(t)=\displaystyle\frac{1}{\mu^2}\big((\rho^2-\Omega^2
)C_3+2\rho\Omega C_4\big)\sin2\Omega t+
\frac{1}{\mu^2}\big((\rho^2-\Omega^2)C_4 -2\rho\Omega C_3\big)
\cos2\Omega t+ \frac{\tilde{\sigma}}{\mu}C_5.\nonumber
\end{eqnarray}
Here $C_l$, $l=\overline{1,5}$ are arbitrary constants.

Following equation (\ref{llst03a30}),  denote  the general
solution of the Hamilton-Ehrenfest system (\ref{spgau2n}),
(\ref{spgau3n}) by  ${\mathfrak g}={\mathfrak g}(t,{\mathfrak
C})\in{\mathbb R}^5$. To solve the spectral problem,  we have need
for the stationary solution of the Hamilton-Ehrenfest system
(\ref{spgau2n}), (\ref{spgau3n}) ${\mathfrak g}={\mathfrak
g}(t,{\mathfrak C}_s)\in{\mathbb R}^5$ that is obtained with the
parameters taken as follows:
\begin{equation}
{\mathfrak C}_s=(0,0,0,0,C_5)^\intercal.
\end{equation}
The system in variations (\ref{gau13q}) for the corresponding
 associated linear equation (\ref{gau6a}) becomes
\begin{equation}
\dot{a}(t)=\left(\begin{array}{cc}
 -\rho &-\widetilde{\sigma} \\
 \mu &\rho \end{array}\right)a(t).\label{vargau12}
\end{equation}
Let us set a Floquet problem \cite{Floke} for the system in
variations (\ref{vargau12}):
\begin{equation}
a(t+T)=e^{i\Omega T}a(t).\label{gau12}
\end{equation}
The quasiperiodicity condition (\ref{gau12}) for solutions of the
system in variations (\ref{vargau12}) is sufficient for the
solutions of the Hamilton-Ehrenfest system in the linear
approximation to be stable. A solution of the Floquet problem
(\ref{vargau12}) and (\ref{gau12}), with the normalization
condition
\[\{a(t), a^*(t)\}=2i, \qquad \{a_1,a_2\}=\langle a_1,J^ta_2\rangle ,\]
can be written in the form
\begin{equation}
a(t)=\frac{e^{i\Omega t}}{\sqrt{\Omega\mu}}\left(\begin{array}{c}
-\rho+i\Omega\\
\mu \end{array}\right).\label{gau13bb}
\end{equation}
Let us seek solutions to equation  (\ref{specgau1}) in the form
(\ref{gau4}). Then the solution (\ref{spgau13a}) of the
correspondent associated equation  (\ref{gau6a}) is given by
\begin{equation}
\Phi_0(x,t,{\mathfrak
C})=\sqrt[4]{\frac{1}{\pi\hbar}}\sqrt[2]{\frac{1}{\mid
C(t)\mid}}\exp{ \Big\{\frac{i}{\hbar}\Big(-\frac{\tilde\varkappa}2
c C_5 t-\frac12\hbar\Omega t+\frac 12\frac{B(t)} {C(t)}\Delta
x^2\Big)\Big\}}. \label{gau13a}
\end{equation}
Accordingly, for the Fock basis  (\ref{gau19}) we obtain
\begin{equation}
\Phi_n^(x,t,{\mathfrak
C})=\frac{i^n}{\sqrt{n!}}\exp\big\{-in\Omega t\big\}
\Big(\frac{1}{\sqrt{2}}\Big)^n H_n\biggl(\sqrt{\frac{\Omega}
{\hbar\mu}}\Delta x\biggr) \Phi_0(x,t,{\mathfrak
C}).\label{gau19q}
\end{equation}
Taking the parameters of the solution of the Hamilton-Ehrenfest
system in the form
\begin{equation}
\overline{\mathfrak C}_n=\big(C_1,C_2,C_3,C_4,C_5\big)^\intercal=
\Big(0,0,0,0,\hbar\frac{\mu(2n+1)}{2\Omega}\Big)^\intercal,
\end{equation}
we obtain that that solutions of the associated equation
(\ref{gau6a}) will be solutions of the original equation
(\ref{specgau1}):
\begin{eqnarray}
&&\Psi_n(x,t)=\Phi_n(x,t,\overline{{\mathfrak C}}_n)=
\frac{i^n}{\sqrt{n!}} \Big(\frac{1}{\sqrt{2}}\Big)^n
\Big(\frac{1}{\pi\hbar}\Big)^{1/4}\Big(\frac{\Omega}{\mu}
\Big)^{1/4}\times\cr &&\quad\times \exp{
\Biggl\{\frac{i}{\hbar}\bigg(-\Big(n+\frac{1}{2}\Big)\hbar\frac{\tilde\varkappa
c\mu}{2\Omega} t-\Big(n+\frac{1}{2}\Big)\hbar\Omega t-\frac
12\frac{\rho} {\mu} x^2\bigg)-\frac {1}{2\hbar}\frac{\Omega} {\mu}
x^2\biggr\}}H_n\biggl(\sqrt{\frac{\Omega} {\hbar\mu}} x\biggr).
\end{eqnarray}
Hence, the eigenfunctions of the operator (\ref{phbb07}) have the
form
\begin{equation}
\psi_n(x,R)=\frac{i^n}{\sqrt{n!}} \Big(\frac{1}{\sqrt{2}}\Big)^n
\Big(\frac{1}{\pi\hbar}\Big)^{1/4}\Big(\frac{\Omega}{\mu}
\Big)^{1/4}\exp\Big\{-\frac{i}{2\hbar}\frac{\rho}{\mu}
x^2-\frac{1}{2\hbar}\frac{\Omega}{\mu} x^2\Big\}
H_n\biggl(\sqrt{\frac{\Omega}{\hbar\mu}}x\biggr),\label{spsob1}
\end{equation}
and the corresponding eigenfunctions are given by
\begin{equation}
E_{n}=\hbar\bigg(n+\frac{1}{2}\bigg)\bigg(\frac{\tilde\varkappa
c\mu}{2\Omega}+\Omega\bigg).
\end{equation}


\section{Adiabatic approximation}

Assume the  evolution of a system is adiabatic. To this case (see
(\ref{phb07})) the parameters of the Hamiltonian  slowly vary in
time, and we can introduce, along with the "fast" time $t$
appearing in the time derivative, a "slow" time $s$ on which the
parameters of the Hamiltonian ($R(t)$=$R(s)$) depend.

Let the  "fast" and the "slow" times be  related as
\begin{equation}
s=\frac tT,\label{adiab01}
\end{equation}
where $T$ is the evolution period of the system.

For  equation (\ref{pgau1}) we consider the Cauchy problem
\begin{eqnarray}
&&\big\{ -i\hbar\partial_t +\widehat{\mathcal
H}_{\varkappa}(R(s),\Psi(t))\big\}\Psi =0,\label{adgau1}\\
&&\Psi(x,t)\big|_{t=0}=\psi_{n}(x,R(0)),\label{adgau1q}
\end{eqnarray}
where $\psi_{n}(x,R(0))$ are the eigenfunctions of the
instantaneous Hamiltonian $\widehat{\mathcal
H}_{\varkappa}(R(0),\Psi(t))$.

As  noted above, to solve  equation (\ref{adgau1}) in adiabatic
approximation we have to solve the Hamilton-Ehrenfest system and
the system in variations accurate to the first order of $1/T $.

The Hamilton-Ehrenfest system for the first-order moments can be
written as
\begin{equation}
\left\{\begin{array}{l}
\displaystyle\frac{1}{T}p'=-\sigma_{0}(s)x-\rho(s) p,\\[6pt]
\displaystyle\frac{1}{T}x' =\mu(s) p+\rho(s) x,
\end{array}\right. \label{adgau3a}
\end{equation}
and for the second-order moments we have
\begin{equation}
\left\{\begin{array}{l}
\displaystyle\frac{1}{T}\sigma'_{xx}=2\mu(s) \sigma_{xp}+2\rho(s)\sigma_{xx}, \\[6pt]
\displaystyle\frac{1}{T}\sigma'_{xp} = \mu(s)
\sigma_{pp}-\widetilde{\sigma}(s)\sigma_{xx},\\[6pt]
\displaystyle\frac{1}{T}\sigma'_{pp}
=-2\rho(s)\sigma_{pp}-2\tilde{\sigma}(s)\sigma_{xp},
\end{array}\right.\label{adgau3}
\end{equation}
where $a'=da/ds$.

Let us seek a solution to the Hamilton-Ehrenfest system
(\ref{adgau3a}),  (\ref{adgau3}) in the form
\begin{eqnarray}
x(t)=x^{(0)}(s)+\frac{1}{T}x^{(1)}(s),\qquad
p(t)=p^{(0)}(s)+\frac{1}{T}p^{(1)}(s), \label{adiab04}\qquad
\Sigma(t)=\Sigma^{(0)}(s)+\frac{1}{T}\Sigma^{(1)}(s),
\end{eqnarray}
where
\begin{equation}
\Sigma=\left(\begin{array}{c}
\sigma_{xx}\\ \sigma_{xp} \\
\sigma_{pp}\end{array}\right).\nonumber
\end{equation}
Substituting  (\ref{adiab04}) in the Hamilton-Ehrenfest system
(\ref{adgau3a}), (\ref{adgau3}) and equating the term of the same
powers in  $1/T$, we obtain
\begin{eqnarray}
&&x^{(0)}=x^{(1)}=0,\qquad p^{(0)}=p^{(1)}=0;\\
&&\Sigma^{(0)}(s)=C_1\left(\begin{array}{c}
\displaystyle\frac{\mu(s)}{\Omega(s)} \\[6pt]
-\displaystyle\frac{\rho(s)}{\Omega(s)}\\[6pt]
\displaystyle\frac{\widetilde{\sigma}(s)}{\Omega(s)}\end{array}\right),\quad
\Sigma^{(1)}(s)=\sigma^{(1)}_{xx}(s)\left(\begin{array}{c}1\\[6pt]
-\displaystyle\frac{\rho(s)}{\mu(s)}\\[6pt]
\displaystyle\frac{\widetilde{\sigma}(s)}{\mu(s)}\end{array}\right)+
C_1\left(\begin{array}{c}0\\[6pt]
\displaystyle\frac{1}{2\rho(s)}\Big(\frac{\mu(s)}{\Omega(s)}\Big)'\\[6pt]
-\displaystyle\frac{1}{\mu(s)}\Big(\frac{\rho(s)}{\Omega(s)}\Big)'
\end{array}\right).\label{adiab05}
\end{eqnarray}
The function  $\sigma_{xx}^{(1)}(s)$ is determined by the
condition of existence of the following approximation and has the
form
\begin{equation}
\sigma_{xx}^{(1)}(s)=C_1\frac{\mu^{2}(s)}{2\Omega^{3}(s)}
\Big(\frac{\rho(s)}{\mu(s)}\Big)'+C_{2}\Big(\frac{\mu(s)}{\Omega(s)}\Big).\label{adiab06}
\end{equation}

A trajectory in the extended phase space  ${\mathfrak
g}={\mathfrak g}(t,{\mathfrak C})\in{\mathbb R}^5$ has the form
\begin{equation}
{\mathfrak g}(t,{\mathfrak C})=\big(0,0,\sigma_{pp}(t,{\mathfrak
C}), \sigma_{px}(t,{\mathfrak C}),\sigma_{xx}(t,{\mathfrak
C})\big)^\intercal, \quad {\mathfrak
C}=\big(C_1,C_2\big)^\intercal.\label{llst03a30qq}
\end{equation}

Applying the change of variables (\ref{adiab01}) to the system in
variations (\ref{gau13q}) we have
\begin{equation}
\frac{1}{T}a'(t)=J{\mathfrak H}_{zz}(t)a(t). \label{adgau13}
\end{equation}
Let us seek a semiclassical asymptotic solution to the system
(\ref{adgau13}) as
\begin{equation}
a(t)=e^{i(T\Phi(s)+\phi(s))}f(t),\label{adiab08}
\end{equation}
\begin{equation}
f(t)=f^{(0)}(s)+\frac{1}{T}f^{(1)}(s).
\end{equation}
Substituting (\ref{adiab08}) in (\ref{adgau13}) and equating the
terms of the same order in $1/T$ we obtain
\begin{eqnarray}
&&\Phi'(s)=\Omega(s),\qquad
\phi'(s)=-\displaystyle\frac{\mu(s)}{2\Omega(s)}\Big(\frac{\rho(s)}{\mu(s)}\Big)',
\cr&&
f^{(0)}(s)=\displaystyle\frac{1}{\sqrt{\Omega(s)\mu(s)}}\left(\begin{array}{c}
-\rho(s)+i\Omega(s)\\
\mu(s) \end{array}\right).\label{gau13bbq}
\end{eqnarray}
Let us expand the vector $f^{(1)}(s)$ in the basis vectors
$f^{(0)}(s)$ and $f^{(0)*}(s)$ to obtain
\begin{equation}
f^{(1)}(s)=\alpha(s) f^{(0)}(s)+\beta(s) f^{(0)*}(s).
\end{equation}
Then
\begin{equation}
\beta(s)=-\frac{\mu(s)}{4\Omega^{2}(s)}\Big(\frac{\rho(s)-i\Omega(s)}{\mu(s)}\Big)'.
\end{equation}
The function $\alpha(s)$ is to be determined from the condition of
existence of the next-order  approximation. We do not give the
explicit form of this approximation, as it does not contribute to
the  terms of the order of $1/T$.

The matrix $Q(t)=B(t)C^{-1}(t)$ accurate to the first order in
$1/T$ reads
\begin{equation}
Q(t)=Q^{(0)}(s)+\frac1{T}Q^{(1)}(s)=
\frac{-\rho(s)+i\Omega(s)}{\mu(s)}+\frac{1}{T}
\frac{-1}{2\Omega(s)}\Big(\frac{\Omega(s)+i\rho(s)}{\mu(s)}\Big)'.\label{adiab11}
\end{equation}
The function
\begin{eqnarray}
&&\Phi_0^{(0)}(x,t,{\mathfrak C})=
\sqrt[4]{\displaystyle\frac{1}{\pi\hbar}}\Big(\displaystyle\frac{1}{\mid
C(t)\mid}\Big)^{1/2}\exp\Big[\frac{i}{2}\Big({\rm Arg}\, C(0)-{\rm
Arg}\, C(t)\Big)\Big]\times\cr&&\quad\times
\exp\Big[\displaystyle\frac{i}{2\hbar}\Big(-\int\limits_0^t
 \tilde\varkappa c(s)\sigma_{xx}(t,{\mathfrak C},\hbar)dt+Q(t)\Delta
 x^{2}\Big)\Bigg],
\end{eqnarray}
is a solution of the associated equation (\ref{gau6a}) accurate to
the first order in $1/T$. Here $\sigma_{xx}(t,{\mathfrak
C},\hbar)$ is determined from expressions (\ref{adiab05}) and
(\ref{adiab06}), and  $Q(t)$ is given by formula (\ref{adiab11}).
For the Fock basis we obtain the following representation:
\begin{eqnarray}
&&\Phi_{n}^{(0)}(x,t,{\mathfrak
C})=\displaystyle\frac{1}{\sqrt{n!}}\Phi_{0}^{(0)}
\frac{1}{\sqrt{2^{n}}}\exp\Big[-in \big({\rm Arg}\, C(t)-{\rm
 Arg}\, C(0)\big)\Big]H_{n}(\xi),\cr
&&\xi=\sqrt{\displaystyle\frac{{\rm Im}\,
Q^{(0)}(s)}{\hbar}}\Delta x.
\end{eqnarray}
In view of  (\ref{adiab11}), we define $\overline{{\mathfrak
C}}_n$ as
\begin{equation}
\overline{{\mathfrak
C}}_n=\Big(\hbar\frac{\mu(2n+1)}{2\Omega},0\Big)^\intercal.
\end{equation}
Accordingly, a solution to equation  (\ref{adgau1}) is determined
as
\begin{equation}
\Psi^{(0)}_n(x,t)=\Phi_n^{(0)}(x,t,\overline{{\mathfrak C}}_n).
\end{equation}
Note that if (\ref{adgau1q}) is taken as  an initial condition for
finding a solution to equation  (\ref{adgau1}), then
\begin{equation}
\Psi(x,t)=\Psi_{n}^{(0)}(x,t)+O\Big(\frac1T\Big),
\end{equation}
where
\begin{equation}
\Psi_{n}^{(0)}(x,t)=\exp\bigg\{-\frac{i}{\hbar}T\int\limits_{0}^{s}E_n(\tau)d\tau+
i\gamma_{n}(s)\bigg\}\psi_{n}(x,R(s)).\label{adiab10}
\end{equation}
The function (\ref{adiab10}) is a solution of equation
(\ref{adgau1}) in the adiabatic approximation. The quantities
\begin{equation}
E_n(s)=\hbar\Big(n+\frac12\Big)\Big(\frac{\tilde\varkappa
c(s)}{2{\rm Im}\, Q^{(0)}(s)}+\Omega(s)\Big)=
\hbar\Big(n+\frac12\Big)\Big(\frac{\tilde\varkappa
c(s)\mu(s)}{2\Omega(s)}+\Omega(s)\Big)
\end{equation}
are the eigenvalues of the instantaneous Hamiltonian
$\hat{\mathcal H}_{\varkappa}(R(s),\psi_n(R(s))$, and
$\gamma_{n}(s)$ has the form
\begin{eqnarray}
&&\gamma_{n}(s)=-\Big(n+\displaystyle\frac12\Big)\int\limits_{0}^{s}
\Big[\phi'(\tau)-\frac{\tilde\varkappa c(\tau)}{2} \frac{{\rm
Im}\, Q^{(1)}(\tau)}{({\rm Im}\, Q^{(0)}(\tau))^2}\Big]d\tau
=-\Big(n+\frac12\Big)\int\limits_{0}^{s}\Big[1-\frac{\tilde\varkappa
c(\tau)}{2}\frac{\mu(\tau)}{\Omega^2(\tau)}\Big]\phi'(\tau)d\tau=\cr
&&=\Big(n+\displaystyle\frac12\Big)\int\limits_{0}^{s}\Big[1-\frac{\tilde\varkappa
c(\tau)}{2}\frac{\mu(\tau)}{\Omega^2(\tau)}\Big]\frac{\mu(\tau)}{2\Omega(\tau)}
\Big(\frac{\rho(\tau)}{\mu(\tau)}\Big)' d\tau,
\quad\Omega(\tau)=\sqrt{[\sigma(\tau)+\tilde\varkappa
a(\tau)]\mu(\tau)-\rho^2(\tau)}.
\end{eqnarray}

Considering the evolution during a period, we obtain
\begin{equation}
\Psi_{n}^{(0)}(x,T)=\exp\bigg\{-\frac{i}{\hbar}T\int\limits_{0}^{1}E_n(s)ds+
i\gamma_{n}(T)\bigg\}\Psi_{n}^{(0)}(x,0).
\end{equation}
Using  (\ref{Ber2_5}) and  (\ref{phb01}), we determine the dynamic
phase
\begin{equation}
\delta_{n}(T)=\Big(n+\frac12\Big)T\int\limits_{0}^{1}\Big(\frac{\tilde\varkappa
c(s)\mu(s)}{2\Omega(s)}+\Omega(s)\Big)ds\label{adiab12}
\end{equation}
and the Berry phase
\begin{equation}
\gamma_{n}(T)=\Big(n+\frac12\Big)\displaystyle\oint\limits_C
\Big[1-\frac{\tilde\varkappa c}{2}\frac{\mu}{\Omega^2(s)}\Big]
\frac{1}{2\Omega }\Big(d\rho-\frac{\rho }{\mu
}d\mu\Big)=\oint\limits_{C}A^{n}_{\mu}d\mu+A^{n}_{\rho}d\rho.
\label{adiab13}
\end{equation}
The components of the "potential"  $A^{n}$ in the parameter space
are determined by the following relations:
\begin{eqnarray}
&&\ A_{\mu}^{n}=-\Big(n+\frac12\Big)\Big[1-\frac{\tilde\varkappa c
}{2}\frac{\mu }{\Omega^2 }\Big] \frac{1}{2\Omega }\frac{\rho }{\mu
},\cr &&\
A_{\rho}^{n}=\Big(n+\frac12\Big)\Big[1-\frac{\tilde\varkappa c
}{2}\frac{\mu }{\Omega^2 }\Big] \frac{1}{2\Omega },\cr &&\
A_{\sigma}^{n}=A_{a}^{n}=A_{b}^{n}=A_{c}^{n}=0.\label{adiab15}
\end{eqnarray}
The Berry phase can be presented in the form of an integral over a
surface $ \Pi $ in the space of parameters, supported by the
contour $C$:
\begin{eqnarray}
&&\gamma_{n}(T)=\displaystyle\Big(n+\frac12\Big)\iint\limits_\Pi
\frac{1}{4\Omega^3}\big\{\bar{\sigma}(d\rho\wedge
d\mu)+\rho(d\mu\wedge d\bar{\sigma})+\mu(d\bar{\sigma}\wedge
d\rho)\big\}-\cr &&\quad-\displaystyle\frac{\tilde\varkappa
c}{2}\frac{3\mu}{4\Omega^5}\big\{\tilde{\sigma}(d\rho\wedge
d\mu)+\rho(d\mu\wedge d\tilde{\sigma})+\mu(d\tilde{\sigma}\wedge
d\rho)\big\},\cr &&\tilde{\sigma}=\sigma+\tilde\varkappa
a,\quad\bar{\sigma}=\sigma+\tilde\varkappa(a+c),\quad\Omega=\sqrt{[\sigma+\tilde\varkappa
a]\mu-\rho^2}.
\end{eqnarray}
Expression (\ref {adiab10}) confirms the statement of the
adiabatic theorem for a nonlinear equation with the Hamiltonian
(\ref {phbb07}) in the  class of trajectory-concentrated
functions.

The difference between the Berry phases for the nonlinear equation
(\ref{pgau1}) and for the linear Schr\'odinger equation
($\tilde\varkappa=0$) is the variation of the frequency $\Omega $
and in the appearing  additional summand
\begin{equation}
-\Big(n+\frac12\Big)\oint\limits_C \frac{\tilde\varkappa c
}{2}\frac{\mu }{\Omega^2 } \frac{1}{2\Omega }
\Big(d\rho-\frac{\rho }{\mu }d\mu\Big).\label{adiab16}
\end{equation}

Note that the calculation of the Berry phase  from the states
(\ref{spsob1}) by  formula (\ref{phb04}) does not give in the term
(\ref{adiab16}).
Accordingly, expression (\ref{adiab15}) for the "potentials"
differs from that  obtained by formula (\ref{phb05}) by terms
proportional to $\tilde\varkappa c $. This is so because the
states (\ref{spsob1}) do not satisfy the superposition principle.
In the limiting case as $\tilde\varkappa\rightarrow 0$, the Berry
phase coincides with the one obtained earlier in \cite{Berry85}
(see also \cite{DodKlMan}).

The classical analog  to  a  Berry phase is known as a Hannay
angle (see, e. g., \cite{Hannay1}), a component in the dynamic
part of the "angle" variable that arises in an adiabatically
evolable integrable dynamical system described in terms of the
"action -- angle" variables. Hannay's angle has a nature similar
to that of Berry's phase, and the relationship between Berry's
phase $\gamma$ of a quantum system and Hannay's angle $\Theta$  of
the corresponding classical system is given by the formula
\begin{equation}
\Theta=-\hbar\frac{\partial\gamma}{\partial
I}=-\frac{\partial\gamma}{\partial n}, \label{adgauh}
\end{equation}
where $I$ is the quantized action and $n$ are quantum numbers. The
differentiation with respect to $n$ is performed as if it would be
a continuous parameter. In a nonlinear case, it is natural to
associate  the Hamiltonian $\hat{\mathcal H}_{\varkappa}$ with an
analog of Hannay's angle by the formula (\ref{adgauh})
\begin{equation}
\Theta_{\varkappa}=-\oint\limits_C \Big[1-\frac{\tilde\varkappa c
}{2}\frac{\mu }{\Omega^2 }\Big]\frac{1}{2\Omega }
\Big(d\rho-\frac{\rho d\mu}{\mu }\Big).
\end{equation}
In the limit $\tilde\varkappa\rightarrow0$, we obtain the well-
known Hannay angle for the generalized harmonic oscillator
\cite{Hannay1}.

\bigskip

\subsection{Conclusions}

%
%

In this paper we have constructed a solution for the one-
dimensional nonstationary Gross-Pitaevskii equation (\ref{pgau1})
and have found explicit expressions for the adiabatic Berry phase.
The eigenfunctions of the instantaneous nonlinear Hamiltonian
(\ref{phbb07}) have been constructed by a semiclassical method
based on the Maslov complex germ theory \cite{Maslov2,BeD2}. This
method, approximate in general, gives exact solutions for the
spectral problem in the case under consideration.

A classical analog of the Berry phase is the Hannay angle
\cite{Hannay1}. We have defined the Hannay angle in terms of
quantum mechanics, since the nonlinear problem requires a special
study of the "classical equations"\, corresponding to the
nonlinear "quantum"\, Gross-Pitaevskii equation. In our
consideration the role of these classical equations is played by
the Hamilton-Ehrenfest system (\ref{gau1b}), which has no
Hamiltonian form relative to the standard Poisson bracket.

The geometric potentials obtained (\ref{adiab15}), which determine
the Berry phase, are Abelian. In the linear quantum mechanical
case ($\varkappa =0$ in equation (\ref{pgau1})) they can be
treated as effective potentials of an electromagnetic field. Such
a situation takes place, e.g., in the theory of the molecular
Aharonov-Bohm effect in the Born-Oppenheimer adiabatic
approximation. The corrections arising due to this effect are
experimentally established in chemical physics \cite{Biswas}.
These potentials were also considered in \cite{Stern}.

In conclusion, we note that the non-Abelian Berry phase is of
interest in quantum computations \cite{Rasetti,Jhonatan}. Such a
phase appears for an instantaneous Hamiltonian having a degenerate
spectrum, which arises in a multidimensional case. The approach
developed in this paper admits mul\-ti\-di\-men\-sional
generalization and the results obtained can be used to study the
relevant problems.

\subsection*{Acknowledgements}
The work was supported in part by  President of the Russian
Federation, Grant  No NSh-1743.2003.2,  Litvinets~F.  was
supported in part by the scholarship of the nonprofit Dynasty
Foundation.

\end{document}